\documentclass[aoas,MSNbibl,nameyear,seceqn,dvips]{arximspdf}
\usepackage{dcolumn}
\usepackage{graphicx}

\doi{10.1214/13-AOAS642} 
\volume{7}
\issue{3}
\pubyear{2013}
\firstpage{1763}
\lastpage{1777}

\makeatletter

\newcolumntype{d}[1]{D{.}{.}{#1}}

\newcommand{\NO}{\mathrm{NO}}

\makeatother

\begin{document}
\begin{frontmatter}

\title{Estimating daily nitrogen dioxide level: Exploring traffic effects}
\runtitle{Estimating daily nitrogen dioxide level}

\begin{aug}
\author[A]{\fnms{Lixun}~\snm{Zhang}\corref{}\thanksref{t1}\ead[label=e1]{lixun.zhang@aya.yale.edu}},
\author[B]{\fnms{Yongtao}~\snm{Guan}\thanksref{t2}\ead[label=e2]{yguan@bus.miami.edu}},
\author[C]{\fnms{Brian~P.}~\snm{Leaderer}\thanksref{t1}\ead[label=e3]{brian.leaderer@yale.edu}}
\and
\author[D]{\fnms{Theodore~R.}~\snm{Holford}\thanksref{t1}\ead[label=e4]{theodore.holford@yale.edu}}
\runauthor{Zhang, Guan, Leaderer and Holford}
\affiliation{Yale University}
\address[A]{L. Zhang\\
State Street Corporation \\
SFC 1507 \\
1 Lincoln St \\
Boston, Massachusetts 02111\\
USA \\
\printead{e1}}
\address[B]{Y. Guan\\
Department of Management Science\\
University of Miami\\
Coral Gables, Florida 33124-6544\\
USA\\
\printead{e2}}
\address[C]{B. P. Leaderer\\
Division of Environmental Health Sciences\\
Yale School of Public Health\\
New Haven, Connecticut 06520\\
USA \\
\printead{e3}}
\address[D]{T. R. Holford\\
Division of Biostatistics\\
Yale School of Public Health\\
New Haven, Connecticut 06520\hspace*{20.85pt}\\
USA \\
\printead{e4}} 
\end{aug}

\thankstext{t1}{Supported in part by NIH Grants ES017416 and ES005410.}
\thankstext{t2}{Supported in part by NSF Grant DMS-08-45368.}

\received{\smonth{3} \syear{2012}}
\revised{\smonth{2} \syear{2013}}

%
\begin{abstract}
Data used to assess acute health effects from air pollution typically
have good temporal but poor spatial resolution or the opposite. A
modified longitudinal model was developed that sought to improve
resolution in both domains by bringing together data from three sources
to estimate daily levels of nitrogen dioxide ($\NO_2$) at a geographic
location. Monthly $\NO_2$ measurements at 316 sites were made available
by the Study of Traffic, Air quality and Respiratory health (STAR).
Four US Environmental Protection Agency monitoring stations have hourly
measurements of $\NO_2$. Finally, the Connecticut Department of
Transportation provides data on traffic density on major roadways,
a~primary contributor to $\NO_2$ pollution. Inclusion of a traffic
variable improved performance of the model, and it provides a method
for estimating exposure at points that do not have direct measurements
of the outcome. This approach can be used to estimate daily variation
in levels of $\NO_2$ over a region.
\end{abstract}

%

\end{frontmatter}

\section{Introduction}
The relationship between traffic and air pollutants such as $\NO_2$ has
been examined using many different
approaches [e.g., \citet{McConnell2010,Maantay2007}]. Proximity
to traffic has frequently been used as a proxy for
traffic related air pollution exposure in environmental
health [\citet{Jerrett2005,McConnell2006}]. In such studies, the
goal is to determine whether there is a relationship between air
pollution and health outcomes. When direct
measurements of specific pollutant levels are not available, proximity
to roadways and traffic levels are sometimes
used as proxies. In general, $\NO_2$ levels decline with distance from a
highway [\citet{Rodes1981,Gilbert2003,Cape2004,Frati2006}].

While data on proximity to major roads have proven to be a
cost-effective approach in epidemiological studies of traffic exposure,
they do not necessarily account for traffic volume. Inclusion of volume
further improves the quality of traffic exposure
measurement [\citet{Rose2009}]. For instance, \citet
{Gauvin2001} found
that including an index of traffic intensity and proximity in a model,
along with an indicator of gas cooker use in the home, improved the
correlation between model estimates and levels of nitrogen dioxide
measured from a monitor located close to a child's home or school.
Other studies [e.g., \citet
{Venn2000,Carr2002,Brauer2003,Heinrich2005,Ryan2005,Schikowski2005,Cesaroni2008}]
also used
traffic volume to improve the quality of exposure information.

One way to include traffic volume information in a model is to
introduce vehicular counts within a buffer zone, which
\citet{Rose2009} call weighted-road-density. The idea is to
calculate the total (road length $\times$ traffic volume)
for a given circle and divide it by the area, that is, $\frac{\sum
_{i=1}^n{L_iV_i}}{\pi r^2}$, where $L_i$ is the length
of a segment, $V_i$ the traffic volume and $r$ the radius of the
circle. Either actual traffic counts or a road
classification system can be used for $V_i$. The authors found that
actual traffic counts were better at predicting
$\NO_2$ than a simple hierarchical classification of roads. In addition,
weighted road density was found to be a better
predictor than proximity to a major road.

Rose et~al.'s (\citeyear{Rose2009}) method assumed that all roads
within a circle
had the same effect regardless of distance to the
point of interest. \citet{Holford2010} proposed a method that
made use of road density, traffic volume and distance to
roads from points of interest. They were able to estimate a dispersion
function for a pollutant, which improved
estimates of $\NO_2$ over those obtained using only average daily
traffic (ADT: number of vehicles/day) on the closest
highway, ADT on the busiest highway within a buffer and the sum for all
road segments within a buffer.

The underlying framework for the methods reviewed above is land use
regression which uses traffic-related variables as
predictors for $\NO_2$ [e.g., \citet
{Briggs1997,Gilbert2005,Gonzales2005,Ross2006,Jerrett2007,Rosenlund2008,Wheeler2008}].
\citet{Ibarra-Berastegi2003} added a time-varying component to a
model using multiple linear regression
to forecast $\NO_2$ levels up to 8 hours in advance by using current and
past 15 hours meteorology along with traffic
information.

Further methods for assessing intraurban exposure were reviewed by
\citet{Jerrett2005}: (i) statistical
interpolation [\citet{Jerrett2001}], (ii)~line dispersion
models [\citet{Bellander2001}], (iii) integrated
emission-meteoro\-logical models [\citet{Frohn2002}], and (iv)
hybrid models combining personal or household exposure
monitoring with one of the preceding methods [\citet
{Kramer2000,Zmirou2002}], or combining two or more of the preceding
methods with regional monitoring [\citet{Hoek2001}]. \citet
{Rose2009} broke down the alternatives into just two
categories: dispersion-based models and empirical models.

As pointed out by \citet{Jerrett2005}, a disadvantage of
geostatistical interpolation is the limited availability of
monitoring data. This approach requires a reasonably dense network of
sampling sites. Government monitoring data
generally come from a sparse network of stations, giving rise to
systematic errors in estimates at sites far from the
monitoring stations. Increasing the number of monitoring sites can be
helpful but costly, so it has not been used
extensively. Researchers often have to use pollution measurements over
relatively short time periods as a substitute
for the comparatively long periods covered by health histories. This
poses a choice between relying on a government
network that provides temporal detail for a limited number of sites or
on their own more detailed spatial network,
which usually covers a short period of time.

To address the limitations inherent in each source of available data,
\citet{Zhang2011} applied a longitudinal model
that established a relationship between data from US Environmental
Protection Agency (EPA) monitoring sites with daily
or finer temporal resolution and those from the Study of Traffic, Air
quality and Respiratory health in children (STAR)
with monthly resolution. It was assumed that the relationship at the
monthly level held at the daily level, using a
model in which data from EPA sites were used to estimate pollution
information at study sites. This model performed
well as measured by $R^2$ in a simple linear model that used STAR site
observations as the response variable and the
predictions based on EPA measurements as the predictor variable. The
model showed that about 73\% of the variability at
the STAR sites can be explained by the predictions. This article
extends and seeks to improve Zhang's (\citeyear{Zhang2011})
method by including traffic as predictors in the model. A
traffic-related variable can then be used to explain the
spatial variation observed in the random intercept of the longitudinal
model, thus providing a practical way for
estimating the temporal/spatial distribution of $\NO_2$ in a region.

\section{Methods}
\subsection{EPA and STAR data}
\begin{figure}

\includegraphics{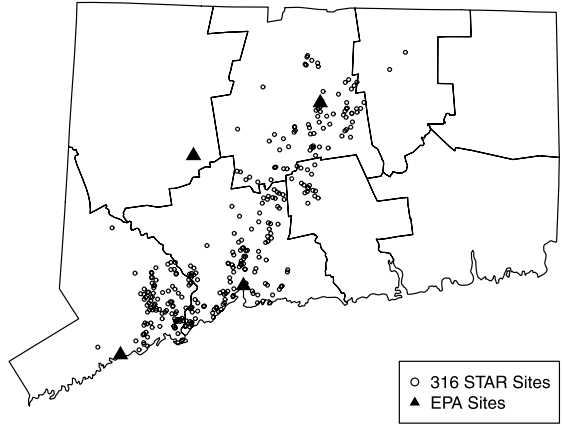}

\caption{Locations of 4 EPA sites which have hourly $\NO_2$
measurements and 316 STAR sites which have monthly measurements.}
\label{locations}
\end{figure}

STAR is an epidemiological study of childhood asthma designed to
investigate whether common air contaminants are
related to disease severity. Four monthly outdoor $\NO_2$ measurements
were taken for each subject, with three months
separating each consecutive measurement. Observations used in this
analysis were taken between April 25, 2006 and March
21, 2008. In contrast to the STAR study, the EPA monitoring sites
provide hourly $\NO_2$ measurements. Average daily
$\NO_2$ was calculated from these hourly measurements. Figure~\ref
{locations} shows the locations of four EPA sites in
Connecticut and 316 STAR study sites used in this analysis. We selected
randomly 266 STAR learning sites for model
development and the remaining 50 sites were used for model validation.

%

\begin{figure}

\includegraphics{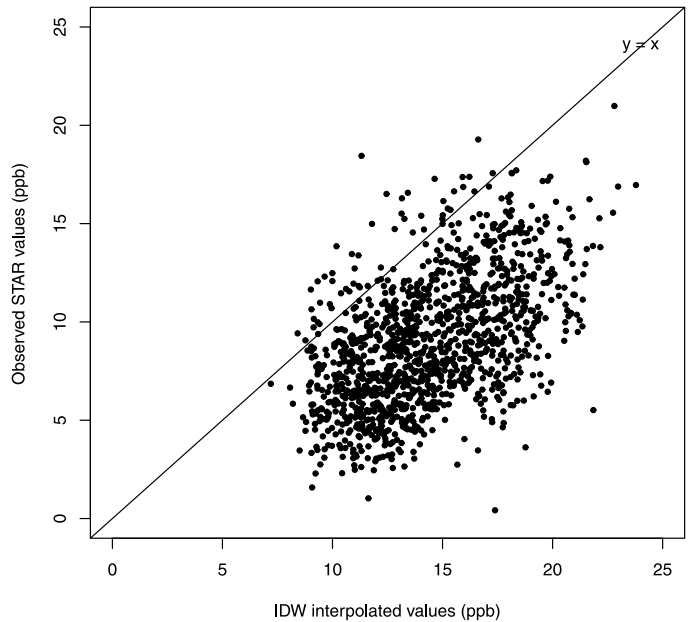}

\caption{Observed $\NO_2$ values at 316 STAR sites ($Z_{i,j}$) vs
average of IDW interpolated values from EPA $\NO_2$
measurements over
the same period as $Z_{i,j}$.}
\label{STAREPAMsr}
\end{figure}

Inverse distance weighting (IDW) was used to interpolate daily $\NO_2$
values at STAR sites based on daily averages at
the four EPA sites. Let $Z_{i,j}$ denote the $j$th $\NO_2$ measurement
at STAR site $i$ (between days $t_1$ to $t_2$,
say), and let $V_{i,t}$ denote the IDW interpolated $\NO_2$ value at
site $i$ on day $t$, for $i=1,2,\ldots,n$, and
$t=1,2,\ldots,T$. A new variable $U_{i,j}$ can be created by taking
the average of $V_{i,t}$ for site $i$ over the same
period as $Z_{i,j}$. Figure~\ref{STAREPAMsr} plots $Z_{i,j}$ against
$U_{i,j}$ for the 316 sites in
Figure~\ref{locations}, where weights are the reciprocal of distance.

%

\subsection{Traffic data} 
\begin{figure}

\includegraphics{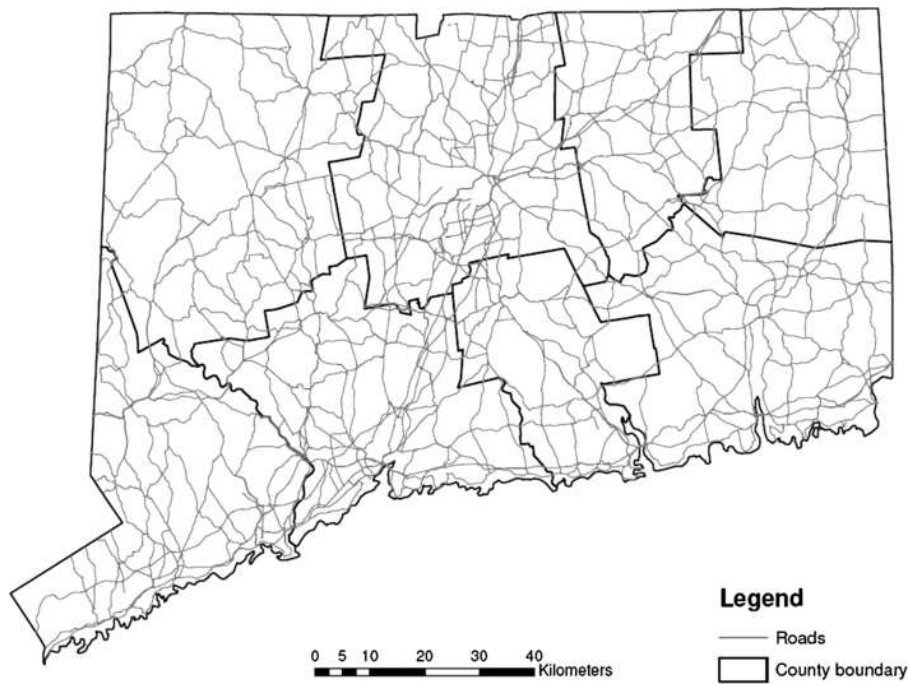}

\caption{Roads with traffic information in Connecticut.}
\label{roadsADT}
\end{figure}

The Connecticut Department of Transportation
reports ADT for all state roads on a three-year cycle. The data for
2006 were used in this analysis. Figure~\ref{roadsADT} shows these
%
%
road segments which have reported
ADT. There are 5196 road segments, with lengths ranging from 16 meters
to 12,295 meters, median of 740 meters and mean
of 1207 meters. The range for ADT was 0 to 184,000 (mean of 22,323 and
median of 11,400).

\subsection{Models}
Three models were compared in this study. First, we considered a linear model:
%
%
\begin{equation}
\label{linearequ} Y_{i} = \alpha_0 + \alpha_1
\times x_{i} + \sum_k
\gamma_k W_{i,k} + \varepsilon_{i},
\end{equation}
where $Y_{i}$ denotes the $i$th $\NO_2$ measurement on the natural log
scale, $x_{i}$ is the natural log of the average
IDW interpolated $\NO_2$ for that site over the corresponding period,
$W_{i,k}$ is the traffic information (ADT), and
$\varepsilon_{i} \sim N(0,\sigma^2)$ is some random error, for $i=1,
2,\ldots, 1064$.

\begin{figure}

\includegraphics{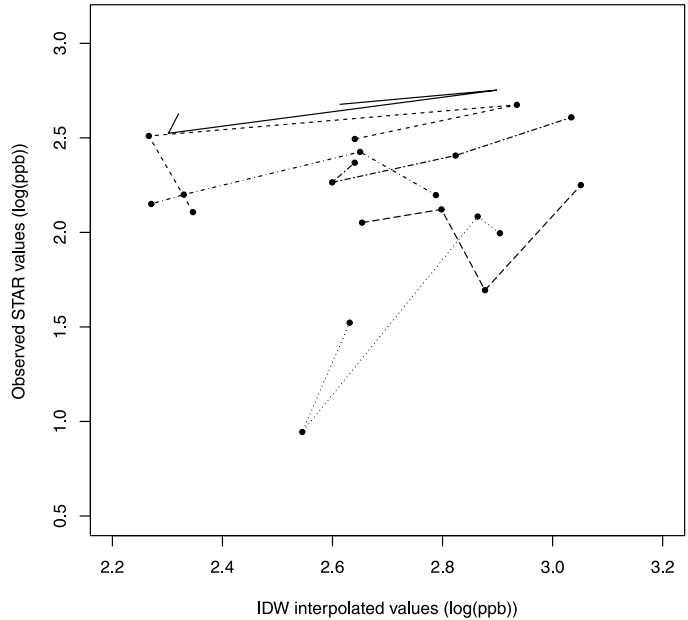}

\caption{Observed $\NO_2$ levels vs averages of IDW interpolated $\NO_2$
levels for six randomly selected STAR sites, with lines connecting
values in temporal order.}
\label{STAR10}
\end{figure}

Second, we specified a longitudinal model with random effects for sites:
%
%
\begin{equation}
\label{longequ} Y_{i,j} = \beta_0 + b_{0,i} +
\beta_1 \times x_{i,j}+ \sum_k
\gamma_k W_{i,k} +\varepsilon_{i,j},
\end{equation}
where $Y_{i,j}$ denotes the $j$th $\NO_2$ measurement at STAR site $i$
on the natural log scale, $x_{i,j}$ is the
corresponding average of IDW interpolated $\NO_2$ on the natural log
scale, $W_{i,k}$ is the traffic information,
$b_{0,i} \sim N(0,\sigma_{b}^2)$ is a random intercept for site $i$,
and $\varepsilon_{i,j} \sim N(0,\sigma_Y^2)$ is some
random error, for $i=1,2,\ldots,266$ and $j=1,2,3,4$. The random
effects $b_{0,i}$ and $\varepsilon_{i,j}$ are mutually
independent. A scatter plot showing this relationship for these data is
shown in Figure~\ref{STAR10}, which shows
$Z_{i,j}$ (the $j$th $\NO_2$ measurement at site $i$) against $U_{i,j}$
(average of IDW interpolated daily $\NO_2$ values
at site $i$ over the period corresponding to $Z_{i,j}$) for six
randomly selected sites, with lines connecting values
for a site in temporal order.

%

Finally, we specified a modified longitudinal model which allowed for
spatial correlation among site effects for the
model in equation (\ref{longequ}), that is, $\mathbf{b_0} = (b_{0,1},
b_{0,2},\ldots, b_{0,n})^T \sim
N(\mathbf{0},\sigma_{b}^2 \times\Sigma(\phi))$. Elements in the
covariance matrix $\Sigma(\phi)$ are given by
$\exp(-\frac{d}{\phi})$, where $d$ denotes spatial distance. The
random effects $\mathbf{b_0}$ and $\varepsilon_{i,j}$'s are
mutually independent.

We adjusted for traffic effects using the integrated exposure model
proposed by \citet{Holford2010} which introduced
covariates into the linear predictor in a regression model. The
contribution of traffic was expressed as
\[
\int z(s) \phi(s) \,ds,
\]
where $z(s)$ denotes ADT for point $s$ on a line representing a highway
and $\phi(s)$ is a dispersion
function for the pollutant generated at $s$. We can achieve
computational efficiency with little loss in accuracy by
representing this contribution numerically---taking the sum of the
product of ADT, the segment length and the unknown
dispersion function which depends on distance. \citet
{Holford2010} discussed alternative forms of linear dispersion
functions, for example, stepped, polynomial or spline. In this example
we used a step function, in which we estimated a value
for the level of dispersion between specified distance intervals,
$D_{k-1}$ and $D_{k}$: $\sum_{j} z_{k,j} \gamma_k
\delta_{k,j} =\gamma_k \sum_{j} z_{k,j} \delta_{k,j}$, where
$\gamma_k$ is the pollution effect from a unit intensity
source within the interval, $z_{k,j}$ is ADT, and $\delta_{k,j}$ is
length of the segment. The linear predictor related
to traffic effects can now be written as
\[
\int z(s) \phi(s) \,ds = \sum_k \biggl(
\gamma_k \sum_{j}z_{k,j}
\delta_{k,j} \biggr) = \sum_k
\gamma_k W_k,
\]
where $W_k = \sum_{j}z_{k,j} \delta_{k,j}$.

ADT is reported in highly variable lengths, and while this approach
might work well for short segments, it can become
problematic for long segments, for example, if the center of one road
is close to a site but most of the remaining segments
are relatively far away. To mitigate this problem, we divided the
segments into smaller subsegments and found that
50-meter segments provided an adequate accuracy. To show this, we
tested lengths such as 10-meter, 50-meter, 100-meter
and up to 5000-meter and found little difference in the resulting
estimates between 10 and 50 meters. For this
example, we used 50-meter. Segments were divided into subsegments using
a Python (\url{http://www.python.org/}) script
which calls relevant ArcGIS [\citet{Institute2010}] functions.

Values of $D_k$'s were predetermined by our experience with earlier
analysis. Setting the values of $D_k$ beforehand
leaves the values of $\gamma_k$'s to be estimated as regression
parameters. Two possible approaches for incorporating
traffic effects were examined: a single-step model which sets the
contribution of highway segments within 2000 meters
as equal and for distances farther than 2000 meters as 0; and a
multi-step model with steps at 400 meters, 800
meters, 1200 meters, 1600 meters and 2000 meters.

While models (\ref{linearequ}) and (\ref{longequ}) were fitted using
a frequentist approach, we obtained parameter
estimates for the third model under the Bayesian framework.

The three models were fitted to $\NO_2$ levels at the 266 learning sites
and the results were used to estimate levels
not only at these sites but at the 50 validation sites as well. By
assuming that the relationship at the monthly level
also holds at the daily level, we also obtained daily estimates. One
predictor variable was based on daily pollution
levels obtained by interpolating with IDW measurements from the four
EPA sites. We also included the remaining
predictors representing traffic-related effects $W_{i,k}$.

Once daily $\NO_2$ predictions at the sites were obtained, they were
averaged over the same periods as the STAR
observations. Systematic departures for site estimates were evaluated
using simple linear regression:
%
%
\begin{equation}
\label{valmodel} Z_{i,j} = \alpha_0 + \alpha_1
* P_{ij}+ \varepsilon_{i,j},
\end{equation}
where $Z_{i,j}$ is the $j$th observation at STAR site $i$, $P_{i,j}$ is
the average of the estimated daily $\NO_2$
values at site $i$ over the same period as $Z_{i,j}$, and $\varepsilon
_{i,j} \sim N(0,\sigma^2)$. In addition, we
calculated the root mean square error (RMSE):
\[
\sqrt{\frac{\Sigma_{i=1}^{n}\Sigma_{j=1}^{4}(Z_{i,j}-P_{i,j})^2}{4n}}.
\]

\section{Results}
Table~\ref{simplelinear} shows results from fitting the model in
equation (\ref{linearequ}) using the single-step and
multi-step dispersion models for the traffic effect. Table~\ref
{longtraffic} shows results from fitting the
corresponding longitudinal model in equation (\ref{longequ}). In
Table~\ref{simplelinear}, the results from the
multi-step dispersion model reveal that the effects of the first two
steps (0--400 m and 400--800 m) are not significantly
different from zero at the 0.05 significance level. While parameter
%
%
\begin{table}[b]
\caption{Results from fitting the linear model in (\protect\ref
{linearequ}) with different traffic variables}\label{simplelinear}%
\begin{tabular*}{\tablewidth}{@{\extracolsep{\fill}}lcd{2.4}cd{2.4}d{2.4}c@{}}
\hline
\textbf{Traffic}& & \multicolumn{1}{c}{\textbf{Estimate}} &
\multicolumn{1}{c}{\textbf{SE}} & \multicolumn{1}{c}{$\bolds{t}$\textbf{-value}}
& \multicolumn{1}{c}{$\bolds{p}$\textbf{-value}} &
\multicolumn{1}{c@{}}{\textbf{Adjusted} $\bolds{R^2}$} \\
\hline
Single-step & $\alpha_0$ & -0.3728 & 0.1181 &
-3.1570 & 0.0016 & 0.3857 \\
& $\alpha_1$ & 0.9428 & 0.0447 & 21.0930 & \mbox{$<$}0.0001 & \\
& $\gamma$ & 0.1524 & 0.0098 & 15.5110 & \mbox{$<$}0.0001 & \\[4pt]
Multi-step & $\alpha_0$ & -0.3963 & 0.1184 & -3.3470
& 0.0008 & 0.3911 \\
& $\alpha_1$ & 0.9341 & 0.0446 & 20.9230 & \mbox{$<$}0.0001 & \\
& $\gamma_1$ & -0.0133 & 0.0283 & -0.4710 & 0.6378 & \\
& $\gamma_2$ & 0.0062 & 0.0236 & 0.2630 & 0.7926 & \\
& $\gamma_3$ & 0.0622 & 0.0233 & 2.6660 & 0.0078 & \\
& $\gamma_4$ & 0.0675 & 0.0151 & 4.4810 & \mbox{$<$}0.0001 & \\
& $\gamma_5$ & 0.0495 & 0.0099 & 4.9900 & \mbox{$<$}0.0001 & \\
\hline
\end{tabular*}
\end{table}
estimates of the next three steps (800--1200 m, 1200--1600 m and
1600--2000 m) are significantly different from zero, their values are
nearly the same (0.0622, 0.0675 and 0.0495). Similar observations can
be made on the results from the longitudinal model in
Table~\ref{longtraffic}. While one might expect values to decline with
distance, this could be due to the high correlation among traffic
covariates for the five steps. The variance inflation factor (VIF) for
each traffic variable in model (\ref{linearequ}) was above one and the
VIFs for two of them were above three. While multi-collinearity does
not greatly affect prediction severely in general, it can be difficult
to diagnose the potential issue of extrapolation with multiple
predictors when making a prediction at a new site. Moreover, note from
Table~\ref{simplelinear} that the adjusted $R^2$ only improved
marginally with the use of multi-step variables. For these reasons we
focused on the model using the single-step traffic variable.

%
%
\begin{table}
\tablewidth=301pt
\caption{Results from fitting the longitudinal model in
(\protect\ref{longequ}) with different traffic variables}
\label{longtraffic}
\begin{tabular*}{\tablewidth}{@{\extracolsep{\fill}}lcd{2.4}ccd{2.4}d{2.4}@{}}
\hline
\textbf{Traffic}& & \multicolumn{1}{c}{\textbf{Estimate}}
& \multicolumn{1}{c}{\textbf{SE}} & \multicolumn{1}{c}{\textbf{DF}}
& \multicolumn{1}{c}{$\bolds{t}$\textbf{-value}} &
\multicolumn{1}{c@{}}{$\bolds{p}$\textbf{-value}} \\
\hline
Single-step & $\beta_0$ & -0.5974 & 0.1033 & 797 &
-5.7826 & \mbox{$<$}0.0001 \\
&$\beta_1$ & 1.0281 & 0.0389 & 797 & 26.4628 & \mbox{$<$}0.0001 \\
&$\gamma$ & 0.1529 & 0.0146 & 264 & 10.5075 & \mbox{$<$}0.0001 \\
&$\sigma_b^2$ & 0.0402 & & & & \\[2pt]
&$\sigma_Y^2$ & 0.0619 & & & & \\[4pt]
Multi-step & $\beta_0$ & -0.6344 & 0.1053 & 797 &
-6.0222 & \mbox{$<$}0.0001 \\
&$\beta_1$ & 1.0250 & 0.0389 & 797 & 26.3591 & \mbox{$<$}0.0001 \\
&$\gamma_1$ & -0.0117 & 0.0419 & 260 & -0.2797 & 0.7800 \\
&$\gamma_2$ & 0.0070 & 0.0350 & 260 & 0.1985 & 0.8428 \\
&$\gamma_3$ & 0.0627 & 0.0346 & 260 & 1.8149 & 0.0707 \\
&$\gamma_4$ & 0.0653 & 0.0223 & 260 & 2.9300 & 0.0037 \\
&$\gamma_5$ & 0.0503 & 0.0147 & 260 & 3.4294 & 0.0007 \\[1pt]
&$\sigma_b^2$ & 0.0398 & & & & \\[2pt]
&$\sigma_Y^2$ & 0.0619 & & & & \\
\hline
\end{tabular*}
\end{table}
%

%
%
\begin{table}[b]
\tablewidth=301pt
\caption{Results from fitting the modified longitudinal model that
includes spatial correlation in (\protect\ref{longequ}) with a
single-step traffic variable}\label{SpatialModel}%
\begin{tabular*}{\tablewidth}{@{\extracolsep{\fill}}ld{2.4}cd{2.4}d{2.4}d{2.4}@{}}
\hline
& \multicolumn{1}{c}{\textbf{Mean}} & \multicolumn{1}{c}{\textbf{SE}} & \multicolumn{1}{c}{\textbf{2.50\%}} & \multicolumn{1}{c}{\textbf{50\%}} &
\multicolumn{1}{c@{}}{\textbf{97.50\%}} \\
\hline
$\beta_0$ & -0.8524 & 0.0896 & -0.9838 & -0.8748 & -0.6251 \\
$\beta_1$ & 1.0828 & 0.0312 & 1.0068 & 1.0828 & 1.1365 \\
$\gamma$ & 0.1023 & 0.0153 & 0.0725 & 0.1023 & 0.1333 \\
$\sigma_b^2$ & 0.0748 & 0.0203 & 0.0419 & 0.0722 & 0.1207 \\
$\sigma_Y^2$ & 0.0648 & 0.0033 & 0.0588 & 0.0647 & 0.0716 \\
$\phi$ & 12.3184 & 3.6682 & 6.5307 & 12.2449 & 19.5918 \\
\hline
\end{tabular*}
\end{table}

The single-step dispersion function was also used for the modified
longitudinal model and the results are shown in
Table~\ref{SpatialModel}. Table~\ref{predSpatialModel} summarizes
results from a comparison of the fitted and the
observed levels at the 50 validation sites using the model in
equation (\ref{valmodel}). Also included are a
comparison of results for models with and without the traffic variable.
Including the traffic variable improved
performance of both the linear and the longitudinal models. For
instance, the predictive $R^2$ for
model (\ref{valmodel}) changed from 0.2617 to 0.4375 and RMSE from
2.9527 to 2.5763 after including traffic variable
in the longitudinal model. The additive bias $\alpha_0$ in the
longitudinal model changed from 1.0821 ($p$-value 0.283)
to 1.2584 ($p$-value 0.0637).

%

%
%
\begin{table}
\caption{Results from a comparison of predicted and observed values
for the 50 validation sites}\label{predSpatialModel}%
\begin{tabular*}{\tablewidth}{@{\extracolsep{\fill}}l c c c d{2.3} d{2.4} c c c@{}}
\hline
& & \multicolumn{1}{c}{\textbf{Estimate}} & \multicolumn{1}{c}{\textbf{SE}}
& \multicolumn{1}{c}{$\bolds{t}$\textbf{-value}} &
\multicolumn{1}{c}{$\bolds{p}$\textbf{-value}} &
\multicolumn{1}{c}{\textbf{Predictive} $\bolds{R^2}$} & \multicolumn{1}{c}{\textbf{RMSE}} &
\multicolumn{1}{c@{}}{\textbf{Traffic}}\\
\hline
Linear &$\alpha_0$ & 0.1163 & 1.1220 & 0.104 & 0.9180 &
0.2605 & 2.9687 & N\\
model &$\alpha_1$ & 1.0526 & 0.1260 & 8.352 & \mbox{$<$}0.0001 & & \\
&$\alpha_0$ & 0.8978 & 0.7073 & 1.269 & 0.206 &
0.4342 & 2.5843 & Y\\
&$\alpha_1$ & 0.9468 & 0.0768 & 12.327 & \mbox{$<$}0.0001 & & \\
[4pt]
Longi- &$\alpha_0$ & 1.0821 & 1.0057 & 1.076 & 0.2830 &
0.2617 & 2.9527 & N\\
tudinal &$\alpha_1$ & 0.9333 & 0.1114 & 8.377 & \mbox{$<$}0.0001 & &\\
model &$\alpha_0$ & 1.2584 & 0.6748 & 1.865 & 0.0637 &
0.4375 & 2.5763 & Y\\
&$\alpha_1$ & 0.8998 & 0.0725 & 12.409 & \mbox{$<$}0.0001 & &\\
[4pt]
Modified & $\alpha_0$ & 0.5247 & 0.5539 & 0.947 & 0.3450 &
0.5807 & 2.2081 & N\\
longi- & $\alpha_1$ & 0.9703 & 0.0586 & 16.560 & \mbox{$<$}0.0001 & & \\
tudinal & $\alpha_0$ & 0.6802 & 0.5131 & 1.326 & 0.1860 &
0.6106 & 2.1311 & Y\\
model & $\alpha_1$ & 0.9527 & 0.0541 & 17.622 & \mbox{$<$}0.0001 & & \\
\hline
\end{tabular*}
\end{table}

\begin{figure}

\includegraphics{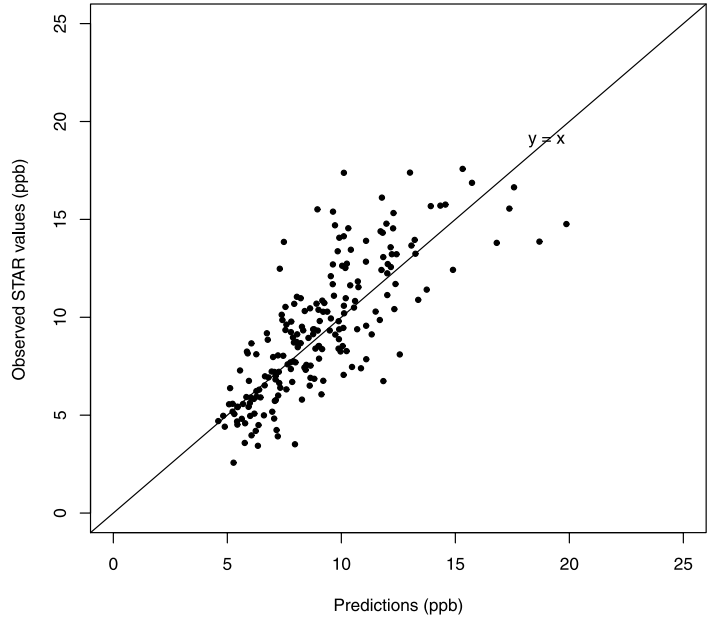}

\caption{Observed vs predicted $\NO_2$ values at 50 validation STAR sites.}
\label{predEst50}
\end{figure}

For the modified longitudinal model that included spatial correlation,
the estimated $\alpha_0$ was not significantly
different from zero, thus being similar to the estimates from the model
without the traffic variable. However, when the
traffic variable was included in this model, the predictive $R^2$ was
0.6106, which was slightly higher than 0.5807 for
the model without traffic. Comparing RMSEs led to similar conclusions,
that is, the model that included traffic had a
lower RMSE compared with the model without traffic. Figure~\ref
{predEst50} shows a scatter plot of observed vs
predicted $\NO_2$ from the modified longitudinal model with traffic effects.

%

%
\begin{figure}

\includegraphics{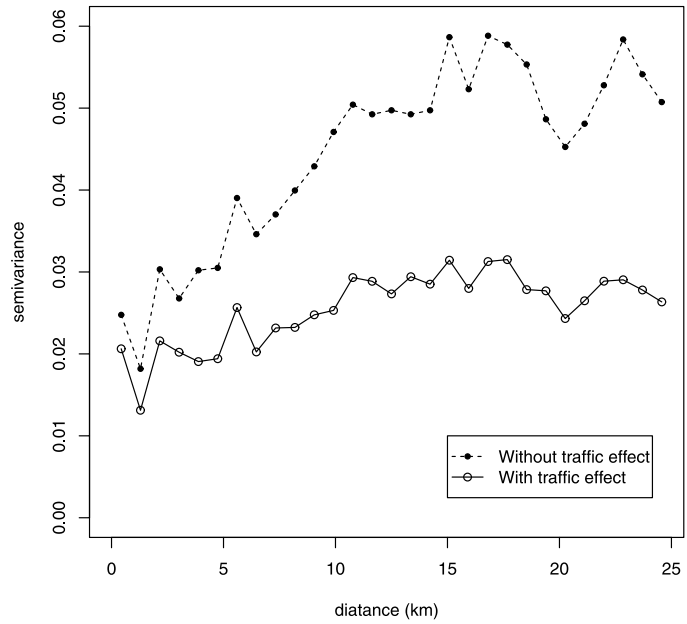}

\caption{Semivariograms of the random intercept in the longitudinal
model, before and after including the predictor for traffic effect.}
\label{semivariogram}
\end{figure}

To see whether traffic effects explain the spatial correlation in the
random intercepts of the longitudinal model, we
compared the sample semivariograms for two versions of the longitudinal
model (\ref{longequ}), one with traffic and
the other without (Figure~\ref{semivariogram}). We can see that the
semivariogram after accounting for traffic is
almost flat compared with the one without traffic. This suggests that
the spatial correlation in the random intercept
has been partially explained by the inclusion of traffic in the model.

\section{Discussion}
Based on the estimated $\alpha_0$, predictive $R^2$ and RMSE for the
50 validation sites, we concluded that inclusion
of traffic effects improved the linear, the longitudinal and the
modified longitudinal models. In addition, the
modified longitudinal model worked reasonably well for making
predictions at random sites.

In the modified longitudinal model, no temporal correlation structure
was assumed for the residual $\varepsilon_{i,j}$. An
area for future research would be to develop a model that allows for
both spatial and temporal correlation.
\citet{Brown2001} and \citet{Romanowicz2006} demonstrated
how such models could be estimated. From an application
perspective, however, assuming only spatial correlation has the
advantage of being less computationally demanding. One
would need to weigh the benefits and costs of using a more complex
model that includes a spatiotemporal correlation
structure.

Another area for further research is to allow for additional predictors
such as land use, population density and
elevation similar to that used by \citet{Skene2010}. In addition,
one needs to explore whether these models can be
applied to different temporal resolutions. The EPA sites record $\NO_2$
levels on an hourly basis, so if the level of
pollutant varies with time of day as a subject moves from place to
place, this could have relevant health consequences.

It would also be useful to determine whether the proposed model can be
applied to other pollutants generated by
traffic. The US EPA monitors a variety of relevant pollutants,
including carbon monoxide, ozone, particulate matter
2.5 and sulfur dioxide. Epidemiological studies have been carried out
to explore the relationship between exposure to
these pollutants and health [e.g., \citet
{Bell2006,Islam2008,Son2011}]. If this approach also performs well for these
pollutants, one would be able to study the effect of daily pollution
levels on health.

Finally, it would be interesting to develop alternative models for
estimating the daily pollution levels at multiple
sites, for example, similar to the latent spatial process used
by \citet{Smith2007}. As a result, it would be no longer
necessary to assume that the relationship between monthly EPA measures
and STAR sites would hold at the daily level.
However, implementation of such models would be computationally
expensive, which could pose a significant challenge for
potential users.

\section*{Acknowledgments}

The authors thank an Associate Editor for the many constructive
comments that greatly enhanced our paper. The authors appreciate the
computing services provided by Yale University Biomedical High
Performance Computing Center, which was funded by NIH Grant RR19895.
The authors also thank Dr. Janneane Gent for her valuable input about
the Study of Traffic, Air Quality and Respiratory Health in Children.


%

\printaddresses


\begin{thebibliography}{42}

\bibitem[\protect\citeauthoryear{Bell and Dominici}{2006}]{Bell2006}
%
\begin{barticle}[author]
\bauthor{\bsnm{Bell},~\bfnm{M.}\binits{M.}} \AND
\bauthor{\bsnm{Dominici},~\bfnm{F.}\binits{F.}}
(\byear{2006}).
\btitle{Analysis of threshold effects for short-term exposure to ozone and
increased risk of mortality}.
\bjournal{Epidemiology}
\bvolume{17}
\bpages{S223--S223}.
\bptok{imsref}%
\end{barticle}
%
\endbibitem

\bibitem[\protect\citeauthoryear{Bellander et~al.}{2001}]{Bellander2001}
%
\begin{barticle}[pbm]
\bauthor{\bsnm{Bellander},~\bfnm{T.}\binits{T.}},
\bauthor{\bsnm{Berglind},~\bfnm{N.}\binits{N.}},
\bauthor{\bsnm{Gustavsson},~\bfnm{P.}\binits{P.}},
\bauthor{\bsnm{Jonson},~\bfnm{T.}\binits{T.}},
\bauthor{\bsnm{Nyberg},~\bfnm{F.}\binits{F.}},
\bauthor{\bsnm{Pershagen},~\bfnm{G.}\binits{G.}} \AND
\bauthor{\bsnm{Jarup},~\bfnm{L.}\binits{L.}}
(\byear{2001}).
\btitle{Using geographic information systems to assess individual historical
exposure to air pollution from traffic and house heating in Stockholm}.
\bjournal{Environ. Health Perspect.}
\bvolume{109}
\bpages{633--639}.
\bid{issn={0091-6765}, pii={sc271_5_1835}, pmcid={1240347}, pmid={11445519}}
\bptok{imsref}%
\end{barticle}
%
\endbibitem

\bibitem[\protect\citeauthoryear{Brauer et~al.}{2003}]{Brauer2003}
%
\begin{barticle}[pbm]
\bauthor{\bsnm{Brauer},~\bfnm{Michael}\binits{M.}},
\bauthor{\bsnm{Hoek},~\bfnm{Gerard}\binits{G.}}, \bauthor{\bparticle{van}
\bsnm{Vliet},~\bfnm{Patricia}\binits{P.}},
\bauthor{\bsnm{Meliefste},~\bfnm{Kees}\binits{K.}},
\bauthor{\bsnm{Fischer},~\bfnm{Paul}\binits{P.}},
\bauthor{\bsnm{Gehring},~\bfnm{Ulrike}\binits{U.}},
\bauthor{\bsnm{Heinrich},~\bfnm{Joachim}\binits{J.}},
\bauthor{\bsnm{Cyrys},~\bfnm{Josef}\binits{J.}},
\bauthor{\bsnm{Bellander},~\bfnm{Tom}\binits{T.}},
\bauthor{\bsnm{Lewne},~\bfnm{Marie}\binits{M.}} \AND
\bauthor{\bsnm{Brunekreef},~\bfnm{Bert}\binits{B.}}
(\byear{2003}).
\btitle{Estimating long-term average particulate air pollution concentrations:
Application of traffic indicators and geographic information systems}.
\bjournal{Epidemiology}
\bvolume{14}
\bpages{228--239}.
\bid{doi={10.1097/01.EDE.0000041910.49046.9B}, issn={1044-3983},
pmid={12606891}}
\bptok{imsref}%
\end{barticle}
%
\endbibitem

\bibitem[\protect\citeauthoryear{Briggs et~al.}{1997}]{Briggs1997}
%
\begin{barticle}[author]
\bauthor{\bsnm{Briggs},~\bfnm{D.~J.}\binits{D.~J.}},
\bauthor{\bsnm{Collins},~\bfnm{S.}\binits{S.}},
\bauthor{\bsnm{Elliott},~\bfnm{P.}\binits{P.}},
\bauthor{\bsnm{Fischer},~\bfnm{P.}\binits{P.}},
\bauthor{\bsnm{Kingham},~\bfnm{S.}\binits{S.}},
\bauthor{\bsnm{Lebret},~\bfnm{E.}\binits{E.}},
\bauthor{\bsnm{Pryl},~\bfnm{K.}\binits{K.}},
\bauthor{\bsnm{VanReeuwijk},~\bfnm{H.}\binits{H.}},
\bauthor{\bsnm{Smallbone},~\bfnm{K.}\binits{K.}} \AND
\bauthor{\bsnm{VanderVeen},~\bfnm{A.}\binits{A.}}
(\byear{1997}).
\btitle{Mapping urban air pollution using GIS: A regression-based approach}.
\bjournal{International Journal of Geographical Information Science}
\bvolume{11}
\bpages{699--718}.
\bptok{imsref}%
\end{barticle}
%
\endbibitem

\bibitem[\protect\citeauthoryear{Brown et~al.}{2001}]{Brown2001}
%
\begin{barticle}[mr]
\bauthor{\bsnm{Brown},~\bfnm{Patrick~E.}\binits{P.~E.}},
\bauthor{\bsnm{Diggle},~\bfnm{Peter~J.}\binits{P.~J.}},
\bauthor{\bsnm{Lord},~\bfnm{Martin~E.}\binits{M.~E.}} \AND
\bauthor{\bsnm{Young},~\bfnm{Peter~C.}\binits{P.~C.}}
(\byear{2001}).
\btitle{Space-time calibration of radar rainfall data}.
\bjournal{J. R. Stat. Soc. Ser. C. Appl. Stat.}
\bvolume{50}
\bpages{221--241}.
\bid{doi={10.1111/1467-9876.00230}, issn={0035-9254}, mr={1833274}}
\bptok{imsref}%
\end{barticle}
%
\endbibitem

\bibitem[\protect\citeauthoryear{Cape et~al.}{2004}]{Cape2004}
%
\begin{barticle}[author]
\bauthor{\bsnm{Cape},~\bfnm{J.~N.}\binits{J.~N.}},
\bauthor{\bsnm{Tang},~\bfnm{Y.~S.}\binits{Y.~S.}}, \bauthor{\bparticle{van}
\bsnm{Dijk},~\bfnm{N.}\binits{N.}},
\bauthor{\bsnm{Love},~\bfnm{L.}\binits{L.}},
\bauthor{\bsnm{Sutton},~\bfnm{M.~A.}\binits{M.~A.}} \AND
\bauthor{\bsnm{Palmer},~\bfnm{S.~C.~F.}\binits{S.~C.~F.}}
(\byear{2004}).
\btitle{Concentrations of ammonia and nitrogen dioxide at roadside
verges, and
their contribution to nitrogen deposition}.
\bjournal{Environmental Pollution}
\bvolume{132}
\bpages{469--478}.
\bptok{imsref}%
\end{barticle}
%
\endbibitem

\bibitem[\protect\citeauthoryear{Carr et~al.}{2002}]{Carr2002}
%
\begin{barticle}[pbm]
\bauthor{\bsnm{Carr},~\bfnm{David}\binits{D.}}, \bauthor{\bparticle{von}
\bsnm{Ehrenstein},~\bfnm{Ondine}\binits{O.}},
\bauthor{\bsnm{Weiland},~\bfnm{Stephan}\binits{S.}},
\bauthor{\bsnm{Wagner},~\bfnm{Claudia}\binits{C.}},
\bauthor{\bsnm{Wellie},~\bfnm{Oliver}\binits{O.}},
\bauthor{\bsnm{Nicolai},~\bfnm{Thomas}\binits{T.}} \AND
\bauthor{\bparticle{von} \bsnm{Mutius},~\bfnm{Erika}\binits{E.}}
(\byear{2002}).
\btitle{Modeling annual benzene, toluene, NO2, and soot concentrations
on the
basis of road traffic characteristics}.
\bjournal{Environ. Res.}
\bvolume{90}
\bpages{111--118}.
\bid{issn={0013-9351}, pii={S0013935102943938}, pmid={12483801}}
\bptok{imsref}%
\end{barticle}
%
\endbibitem

\bibitem[\protect\citeauthoryear{Cesaroni et~al.}{2008}]{Cesaroni2008}
%
\begin{barticle}[pbm]
\bauthor{\bsnm{Cesaroni},~\bfnm{G.}\binits{G.}},
\bauthor{\bsnm{Badaloni},~\bfnm{C.}\binits{C.}},
\bauthor{\bsnm{Porta},~\bfnm{D.}\binits{D.}},
\bauthor{\bsnm{Forastiere},~\bfnm{F.}\binits{F.}} \AND
\bauthor{\bsnm{Perucci},~\bfnm{C.~A.}\binits{C.~A.}}
(\byear{2008}).
\btitle{Comparison between various indices of exposure to
traffic-related air
pollution and their impact on respiratory health in adults}.
\bjournal{Occup. Environ. Med.}
\bvolume{65}
\bpages{683--690}.
\bid{doi={10.1136/oem.2007.037846}, issn={1470-7926}, pii={oem.2007.037846},
pmcid={2771851}, pmid={18203803}}
\bptok{imsref}%
\end{barticle}
%
\endbibitem

\bibitem[\protect\citeauthoryear{Environmental Systems Resource
Institute}{2010}]{Institute2010}
%
\begin{bmisc}[author]
\borganization{Environmental Systems Resource Institute}
(\byear{2010}).
\bhowpublished{ArcMap 10.0. ESRI, Redlands, CA}.
\bptok{imsref}%
\end{bmisc}
%
\endbibitem

\bibitem[\protect\citeauthoryear{Frati et~al.}{2006}]{Frati2006}
%
\begin{barticle}[pbm]
\bauthor{\bsnm{Frati},~\bfnm{L.}\binits{L.}},
\bauthor{\bsnm{Caprasecca},~\bfnm{E.}\binits{E.}},
\bauthor{\bsnm{Santoni},~\bfnm{S.}\binits{S.}},
\bauthor{\bsnm{Gaggi},~\bfnm{C.}\binits{C.}},
\bauthor{\bsnm{Guttova},~\bfnm{A.}\binits{A.}},
\bauthor{\bsnm{Gaudino},~\bfnm{S.}\binits{S.}},
\bauthor{\bsnm{Pati},~\bfnm{A.}\binits{A.}},
\bauthor{\bsnm{Rosamilia},~\bfnm{S.}\binits{S.}},
\bauthor{\bsnm{Pirintsos},~\bfnm{S.~A.}\binits{S.~A.}} \AND
\bauthor{\bsnm{Loppi},~\bfnm{S.}\binits{S.}}
(\byear{2006}).
\btitle{Effects of NO2 and NH3 from road traffic on epiphytic lichens}.
\bjournal{Environ. Pollut.}
\bvolume{142}
\bpages{58--64}.
\bid{doi={10.1016/j.envpol.2005.09.020}, issn={0269-7491},
pii={S0269-7491(05)00487-2}, pmid={16310300}}
\bptok{imsref}%
\end{barticle}
%
\endbibitem

\bibitem[\protect\citeauthoryear{Frohn, Christensen and
Brandt}{2002}]{Frohn2002}
%
\begin{barticle}[author]
\bauthor{\bsnm{Frohn},~\bfnm{L.~M.}\binits{L.~M.}},
\bauthor{\bsnm{Christensen},~\bfnm{J.~H.}\binits{J.~H.}} \AND
\bauthor{\bsnm{Brandt},~\bfnm{J.}\binits{J.}}
(\byear{2002}).
\btitle{Development of a high-resolution nested air pollution model---The
numerical approach}.
\bjournal{J. Comput. Phys.}
\bvolume{179}
\bpages{68--94}.
\bptok{imsref}%
\end{barticle}
%
\endbibitem

\bibitem[\protect\citeauthoryear{Gauvin et~al.}{2001}]{Gauvin2001}
%
\begin{bmisc}[pbm]
\bauthor{\bsnm{Gauvin},~\bfnm{S.}\binits{S.}},
\bauthor{\bsnm{Moullec},~\bfnm{Y.~Le}\binits{Y.~L.}},
\bauthor{\bsnm{Bremont},~\bfnm{F.}\binits{F.}},
\bauthor{\bsnm{Momas},~\bfnm{I.}\binits{I.}},
\bauthor{\bsnm{Balducci},~\bfnm{F.}\binits{F.}},
\bauthor{\bsnm{Ciognard},~\bfnm{F.}\binits{F.}},
\bauthor{\bsnm{Poilve},~\bfnm{M.~P.}\binits{M.~P.}},
\bauthor{\bsnm{Zmirou},~\bfnm{D.}\binits{D.}} \AND
\borganization{VESTA Investigators}
(\byear{2001}).
\bhowpublished{Relationships between nitrogen dioxide personal exposure and
ambient air monitoring measurements among children in three French
metropolitan areas: VESTA study. \textit{Arch. Environ. Health} \textbf{56}
336--341}.
\bid{doi={10.1080/00039890109604465}, issn={0003-9896}, pmid={11572277}}
\bptok{imsref}%
\end{bmisc}
%
\endbibitem

\bibitem[\protect\citeauthoryear{Gilbert et~al.}{2003}]{Gilbert2003}
%
\begin{barticle}[pbm]
\bauthor{\bsnm{Gilbert},~\bfnm{Nicolas~L.}\binits{N.~L.}},
\bauthor{\bsnm{Woodhouse},~\bfnm{Sandy}\binits{S.}},
\bauthor{\bsnm{Stieb},~\bfnm{David~M.}\binits{D.~M.}} \AND
\bauthor{\bsnm{Brook},~\bfnm{Jeffrey~R.}\binits{J.~R.}}
(\byear{2003}).
\btitle{Ambient nitrogen dioxide and distance from a major highway}.
\bjournal{Sci. Total Environ.}
\bvolume{312}
\bpages{43--46}.
\bid{doi={10.1016/S0048-9697(03)00228-6}, issn={0048-9697},
pii={S0048-9697(03)00228-6}, pmid={12873397}}
\bptok{imsref}%
\end{barticle}
%
\endbibitem

\bibitem[\protect\citeauthoryear{Gilbert et~al.}{2005}]{Gilbert2005}
%
\begin{barticle}[author]
\bauthor{\bsnm{Gilbert},~\bfnm{N.~L.}\binits{N.~L.}},
\bauthor{\bsnm{Goldberg},~\bfnm{M.~S.}\binits{M.~S.}},
\bauthor{\bsnm{Beckerman},~\bfnm{B.}\binits{B.}},
\bauthor{\bsnm{Brook},~\bfnm{J.~R.}\binits{J.~R.}} \AND
\bauthor{\bsnm{Jerrett},~\bfnm{M.}\binits{M.}}
(\byear{2005}).
\btitle{Assessing spatial variability of ambient nitrogen dioxide in Montreal,
Canada, with a land-use regression model}.
\bjournal{Journal of the Air Waste Management Association}
\bvolume{55}
\bpages{1059--1063}.
\bptok{imsref}%
\end{barticle}
%
\endbibitem

\bibitem[\protect\citeauthoryear{Gonzales et~al.}{2005}]{Gonzales2005}
%
\begin{barticle}[author]
\bauthor{\bsnm{Gonzales},~\bfnm{M.}\binits{M.}},
\bauthor{\bsnm{Qualls},~\bfnm{C.}\binits{C.}},
\bauthor{\bsnm{Hudgens},~\bfnm{E.}\binits{E.}} \AND
\bauthor{\bsnm{Neas},~\bfnm{L.}\binits{L.}}
(\byear{2005}).
\btitle{Characterization of a spatial gradient of nitrogen dioxide
across a
United States-Mexico border city during winter}.
\bjournal{Science of the Total Environment}
\bvolume{337}
\bpages{163--173}.
\bptok{imsref}%
\end{barticle}
%
\endbibitem

\bibitem[\protect\citeauthoryear{Heinrich et~al.}{2005}]{Heinrich2005}
%
\begin{barticle}[pbm]
\bauthor{\bsnm{Heinrich},~\bfnm{J.}\binits{J.}},
\bauthor{\bsnm{Gehring},~\bfnm{U.}\binits{U.}},
\bauthor{\bsnm{Cyrys},~\bfnm{J.}\binits{J.}},
\bauthor{\bsnm{Brauer},~\bfnm{M.}\binits{M.}},
\bauthor{\bsnm{Hoek},~\bfnm{G.}\binits{G.}},
\bauthor{\bsnm{Fischer},~\bfnm{P.}\binits{P.}},
\bauthor{\bsnm{Bellander},~\bfnm{T.}\binits{T.}} \AND
\bauthor{\bsnm{Brunekreef},~\bfnm{B.}\binits{B.}}
(\byear{2005}).
\btitle{Exposure to traffic related air pollutants: Self reported traffic
intensity versus GIS modelled exposure}.
\bjournal{Occup. Environ. Med.}
\bvolume{62}
\bpages{517--523}.
\bid{doi={10.1136/oem.2004.016766}, issn={1470-7926}, pii={62/8/517},
pmcid={1741068}, pmid={16046603}}
\bptok{imsref}%
\end{barticle}
%
\endbibitem

\bibitem[\protect\citeauthoryear{Hoek et~al.}{2001}]{Hoek2001}
%
\begin{barticle}[pbm]
\bauthor{\bsnm{Hoek},~\bfnm{G.}\binits{G.}},
\bauthor{\bsnm{Fischer},~\bfnm{P.}\binits{P.}},
\bauthor{\bsnm{Brandt},~\bfnm{P.~Van~Den}\binits{P.~V.~D.}},
\bauthor{\bsnm{Goldbohm},~\bfnm{S.}\binits{S.}} \AND
\bauthor{\bsnm{Brunekreef},~\bfnm{B.}\binits{B.}}
(\byear{2001}).
\btitle{Estimation of long-term average exposure to outdoor air
pollution for a
cohort study on mortality}.
\bjournal{J. Expo. Anal. Environ. Epidemiol.}
\bvolume{11}
\bpages{459--469}.
\bid{doi={10.1038/sj.jea.7500189}, issn={1053-4245}, pmid={11791163}}
\bptok{imsref}%
\end{barticle}
%
\endbibitem

\bibitem[\protect\citeauthoryear{Holford et~al.}{2010}]{Holford2010}
%
\begin{barticle}[mr]
\bauthor{\bsnm{Holford},~\bfnm{Theodore~R.}\binits{T.~R.}},
\bauthor{\bsnm{Ebisu},~\bfnm{Keita}\binits{K.}},
\bauthor{\bsnm{McKay},~\bfnm{Lisa~A.}\binits{L.~A.}},
\bauthor{\bsnm{Gent},~\bfnm{Janneane~F.}\binits{J.~F.}},
\bauthor{\bsnm{Triche},~\bfnm{Elizabeth~W.}\binits{E.~W.}},
\bauthor{\bsnm{Bracken},~\bfnm{Michael~B.}\binits{M.~B.}} \AND
\bauthor{\bsnm{Leaderer},~\bfnm{Brian~P.}\binits{B.~P.}}
(\byear{2010}).
\btitle{Integrated exposure modeling: A model using {GIS} and {GLM}}.
\bjournal{Stat. Med.}
\bvolume{29}
\bpages{116--129}.
\bid{doi={10.1002/sim.3732}, issn={0277-6715}, mr={2751384}}
\bptok{imsref}%
\end{barticle}
%
\endbibitem

\bibitem[\protect\citeauthoryear{Ibarra-Berastegi
et~al.}{2003}]{Ibarra-Berastegi2003}
%
\begin{barticle}[author]
\bauthor{\bsnm{Ibarra-Berastegi},~\bfnm{G.}\binits{G.}},
\bauthor{\bsnm{Madariaga},~\bfnm{I.}\binits{I.}},
\bauthor{\bsnm{Agirre},~\bfnm{E.}\binits{E.}} \AND
\bauthor{\bsnm{Uria},~\bfnm{J.}\binits{J.}}
(\byear{2003}).
\btitle{Short-term forecasting of ozone and NO2 levels using traffic
data in
Bilbao (Spain)}.
\bjournal{Urban Transport Ix: Urban Transport and the Environment in
the 21st
Century}
\bvolume{14}
\bpages{235--242, 709}.
\bptok{imsref}%
\end{barticle}
%
\endbibitem

\bibitem[\protect\citeauthoryear{Islam et~al.}{2008}]{Islam2008}
%
\begin{barticle}[pbm]
\bauthor{\bsnm{Islam},~\bfnm{Talat}\binits{T.}},
\bauthor{\bsnm{McConnell},~\bfnm{Rob}\binits{R.}},
\bauthor{\bsnm{Gauderman},~\bfnm{W.~James}\binits{W.~J.}},
\bauthor{\bsnm{Avol},~\bfnm{Ed}\binits{E.}},
\bauthor{\bsnm{Peters},~\bfnm{John~M.}\binits{J.~M.}} \AND
\bauthor{\bsnm{Gilliland},~\bfnm{Frank~D.}\binits{F.~D.}}
(\byear{2008}).
\btitle{Ozone, oxidant defense genes, and risk of asthma during adolescence}.
\bjournal{Am. J. Respir. Crit. Care Med.}
\bvolume{177}
\bpages{388--395}.
\bid{doi={10.1164/rccm.200706-863OC}, issn={1535-4970}, pii={200706-863OC},
pmcid={2258440}, pmid={18048809}}
\bptok{imsref}%
\end{barticle}
%
\endbibitem

\bibitem[\protect\citeauthoryear{Jerrett et~al.}{2001}]{Jerrett2001}
%
\begin{barticle}[author]
\bauthor{\bsnm{Jerrett},~\bfnm{M.}\binits{M.}},
\bauthor{\bsnm{Burnett},~\bfnm{R.~T.}\binits{R.~T.}},
\bauthor{\bsnm{Kanaroglou},~\bfnm{P.}\binits{P.}},
\bauthor{\bsnm{Eyles},~\bfnm{J.}\binits{J.}},
\bauthor{\bsnm{Finkelstein},~\bfnm{N.}\binits{N.}},
\bauthor{\bsnm{Giovis},~\bfnm{C.}\binits{C.}} \AND
\bauthor{\bsnm{Brook},~\bfnm{J.~R.}\binits{J.~R.}}
(\byear{2001}).
\btitle{A GIS---Environmental justice analysis of particulate air
pollution in
Hamilton, Canada}.
\bjournal{Environment and Planning A}
\bvolume{33}
\bpages{955--973}.
\bptok{imsref}%
\end{barticle}
%
\endbibitem

\bibitem[\protect\citeauthoryear{Jerrett et~al.}{2005}]{Jerrett2005}
%
\begin{barticle}[pbm]
\bauthor{\bsnm{Jerrett},~\bfnm{Michael}\binits{M.}},
\bauthor{\bsnm{Arain},~\bfnm{Altaf}\binits{A.}},
\bauthor{\bsnm{Kanaroglou},~\bfnm{Pavlos}\binits{P.}},
\bauthor{\bsnm{Beckerman},~\bfnm{Bernardo}\binits{B.}},
\bauthor{\bsnm{Potoglou},~\bfnm{Dimitri}\binits{D.}},
\bauthor{\bsnm{Sahsuvaroglu},~\bfnm{Talar}\binits{T.}},
\bauthor{\bsnm{Morrison},~\bfnm{Jason}\binits{J.}} \AND
\bauthor{\bsnm{Giovis},~\bfnm{Chris}\binits{C.}}
(\byear{2005}).
\btitle{A review and evaluation of intraurban air pollution exposure models}.
\bjournal{J. Expo. Anal. Environ. Epidemiol.}
\bvolume{15}
\bpages{185--204}.
\bid{doi={10.1038/sj.jea.7500388}, issn={1053-4245}, pii={7500388},
pmid={15292906}}
\bptok{imsref}%
\end{barticle}
%
\endbibitem

\bibitem[\protect\citeauthoryear{Jerrett et~al.}{2007}]{Jerrett2007}
%
\begin{barticle}[author]
\bauthor{\bsnm{Jerrett},~\bfnm{M.}\binits{M.}},
\bauthor{\bsnm{Arain},~\bfnm{M.~A.}\binits{M.~A.}},
\bauthor{\bsnm{Kanaroglou},~\bfnm{P.}\binits{P.}},
\bauthor{\bsnm{Beckerman},~\bfnm{B.}\binits{B.}},
\bauthor{\bsnm{Crouse},~\bfnm{D.}\binits{D.}},
\bauthor{\bsnm{Gilbert},~\bfnm{N.~L.}\binits{N.~L.}},
\bauthor{\bsnm{Brook},~\bfnm{J.~R.}\binits{J.~R.}},
\bauthor{\bsnm{Finkelstein},~\bfnm{N.}\binits{N.}} \AND
\bauthor{\bsnm{Finkelstein},~\bfnm{M.~M.}\binits{M.~M.}}
(\byear{2007}).
\btitle{Modeling the intraurban variability of ambient traffic
pollution in
Toronto, Canada}.
\bjournal{Journal of Toxicology and Environmental Health-Part A-Current Issues}
\bvolume{70}
\bpages{200--212}.
\bptok{imsref}%
\end{barticle}
%
\endbibitem

\bibitem[\protect\citeauthoryear{Kramer et~al.}{2000}]{Kramer2000}
%
\begin{barticle}[pbm]
\bauthor{\bsnm{Kramer},~\bfnm{U.}\binits{U.}},
\bauthor{\bsnm{Koch},~\bfnm{T.}\binits{T.}},
\bauthor{\bsnm{Ranft},~\bfnm{U.}\binits{U.}},
\bauthor{\bsnm{Ring},~\bfnm{J.}\binits{J.}} \AND
\bauthor{\bsnm{Behrendt},~\bfnm{H.}\binits{H.}}
(\byear{2000}).
\btitle{Traffic-related air pollution is associated with atopy in children
living in urban areas}.
\bjournal{Epidemiology}
\bvolume{11}
\bpages{64--70}.
\bid{issn={1044-3983}, pmid={10615846}}
\bptok{imsref}%
\end{barticle}
%
\endbibitem

\bibitem[\protect\citeauthoryear{Maantay}{2007}]{Maantay2007}
%
\begin{barticle}[pbm]
\bauthor{\bsnm{Maantay},~\bfnm{Juliana}\binits{J.}}
(\byear{2007}).
\btitle{Asthma and air pollution in the Bronx: Methodological and data
considerations in using GIS for environmental justice and health research}.
\bjournal{Health Place}
\bvolume{13}
\bpages{32--56}.
\bid{doi={10.1016/j.healthplace.2005.09.009}, issn={1353-8292},
pii={S1353-8292(05)00067-5}, pmid={16311064}}
\bptok{imsref}%
\end{barticle}
%
\endbibitem

\bibitem[\protect\citeauthoryear{McConnell et~al.}{2006}]{McConnell2006}
%
\begin{barticle}[author]
\bauthor{\bsnm{McConnell},~\bfnm{R.}\binits{R.}},
\bauthor{\bsnm{Berhane},~\bfnm{K.}\binits{K.}},
\bauthor{\bsnm{Yao},~\bfnm{L.}\binits{L.}},
\bauthor{\bsnm{Jerrett},~\bfnm{M.}\binits{M.}},
\bauthor{\bsnm{Lurmann},~\bfnm{F.}\binits{F.}},
\bauthor{\bsnm{Gilliland},~\bfnm{F.}\binits{F.}},
\bauthor{\bsnm{Kunzli},~\bfnm{N.}\binits{N.}},
\bauthor{\bsnm{Gauderman},~\bfnm{J.}\binits{J.}},
\bauthor{\bsnm{Avol},~\bfnm{E.}\binits{E.}},
\bauthor{\bsnm{Thomas},~\bfnm{D.}\binits{D.}} \AND
\bauthor{\bsnm{Peters},~\bfnm{J.}\binits{J.}}
(\byear{2006}).
\btitle{Traffic, susceptibility, and childhood asthma}.
\bjournal{Environmental Health Perspectives}
\bvolume{114}
\bpages{766--772}.
\bptok{imsref}%
\end{barticle}
%
\endbibitem

\bibitem[\protect\citeauthoryear{McConnell et~al.}{2010}]{McConnell2010}
%
\begin{barticle}[pbm]
\bauthor{\bsnm{McConnell},~\bfnm{Rob}\binits{R.}},
\bauthor{\bsnm{Islam},~\bfnm{Talat}\binits{T.}},
\bauthor{\bsnm{Shankardass},~\bfnm{Ketan}\binits{K.}},
\bauthor{\bsnm{Jerrett},~\bfnm{Michael}\binits{M.}},
\bauthor{\bsnm{Lurmann},~\bfnm{Fred}\binits{F.}},
\bauthor{\bsnm{Gilliland},~\bfnm{Frank}\binits{F.}},
\bauthor{\bsnm{Gauderman},~\bfnm{Jim}\binits{J.}},
\bauthor{\bsnm{Avol},~\bfnm{Ed}\binits{E.}},
\bauthor{\bsnm{K{\"{u}}nzli},~\bfnm{Nino}\binits{N.}},
\bauthor{\bsnm{Yao},~\bfnm{Ling}\binits{L.}},
\bauthor{\bsnm{Peters},~\bfnm{John}\binits{J.}} \AND
\bauthor{\bsnm{Berhane},~\bfnm{Kiros}\binits{K.}}
(\byear{2010}).
\btitle{Childhood incident asthma and traffic-related air pollution at
home and
school}.
\bjournal{Environ. Health Perspect.}
\bvolume{118}
\bpages{1021--1026}.
\bid{doi={10.1289/ehp.0901232}, issn={1552-9924}, pmcid={2920902},
pmid={20371422}}
\bptok{imsref}%
\end{barticle}
%
\endbibitem

\bibitem[\protect\citeauthoryear{Rodes and Holland}{1981}]{Rodes1981}
%
\begin{barticle}[author]
\bauthor{\bsnm{Rodes},~\bfnm{C.~E.}\binits{C.~E.}} \AND
\bauthor{\bsnm{Holland},~\bfnm{D.~M.}\binits{D.~M.}}
(\byear{1981}).
\btitle{Variations of No, No2 and O-3 concentrations downwind of a Los-Angeles
freeway}.
\bjournal{Atmospheric Environment}
\bvolume{15}
\bpages{243--250}.
\bptok{imsref}%
\end{barticle}
%
\endbibitem

\bibitem[\protect\citeauthoryear{Romanowicz et~al.}{2006}]{Romanowicz2006}
%
\begin{barticle}[author]
\bauthor{\bsnm{Romanowicz},~\bfnm{R.}\binits{R.}},
\bauthor{\bsnm{Young},~\bfnm{P.}\binits{P.}},
\bauthor{\bsnm{Brown},~\bfnm{P.}\binits{P.}} \AND
\bauthor{\bsnm{Diggle},~\bfnm{P.}\binits{P.}}
(\byear{2006}).
\btitle{A recursive estimation approach to the spatio-temporal analysis and
modelling of air quality data}.
\bjournal{Environmental Modelling Software}
\bvolume{21}
\bpages{759--769}.
\bptok{imsref}%
\end{barticle}
%
\endbibitem

\bibitem[\protect\citeauthoryear{Rose et~al.}{2009}]{Rose2009}
%
\begin{barticle}[author]
\bauthor{\bsnm{Rose},~\bfnm{N.}\binits{N.}},
\bauthor{\bsnm{Cowie},~\bfnm{C.}\binits{C.}},
\bauthor{\bsnm{Gillett},~\bfnm{R.}\binits{R.}} \AND
\bauthor{\bsnm{Marks},~\bfnm{G.~B.}\binits{G.~B.}}
(\byear{2009}).
\btitle{Weighted road density: A~simple way of assigning
traffic-related air
pollution exposure}.
\bjournal{Atmospheric Environment}
\bvolume{43}
\bpages{5009--5014}.
\bptok{imsref}%
\end{barticle}
%
\endbibitem

\bibitem[\protect\citeauthoryear{Rosenlund et~al.}{2008}]{Rosenlund2008}
%
\begin{barticle}[pbm]
\bauthor{\bsnm{Rosenlund},~\bfnm{Mats}\binits{M.}},
\bauthor{\bsnm{Forastiere},~\bfnm{Francesco}\binits{F.}},
\bauthor{\bsnm{Stafoggia},~\bfnm{Massimo}\binits{M.}},
\bauthor{\bsnm{Porta},~\bfnm{Daniela}\binits{D.}},
\bauthor{\bsnm{Perucci},~\bfnm{Mara}\binits{M.}},
\bauthor{\bsnm{Ranzi},~\bfnm{Andrea}\binits{A.}},
\bauthor{\bsnm{Nussio},~\bfnm{Fabio}\binits{F.}} \AND
\bauthor{\bsnm{Perucci},~\bfnm{Carlo~A.}\binits{C.~A.}}
(\byear{2008}).
\btitle{Comparison of regression models with land-use and emissions
data to
predict the spatial distribution of traffic-related air pollution in Rome}.
\bjournal{J. Expo. Sci. Environ. Epidemiol.}
\bvolume{18}
\bpages{192--199}.
\bid{doi={10.1038/sj.jes.7500571}, issn={1559-064X}, pii={7500571},
pmid={17426734}}
\bptok{imsref}%
\end{barticle}
%
\endbibitem

\bibitem[\protect\citeauthoryear{Ross et~al.}{2006}]{Ross2006}
%
\begin{barticle}[pbm]
\bauthor{\bsnm{Ross},~\bfnm{Zev}\binits{Z.}},
\bauthor{\bsnm{English},~\bfnm{Paul~B.}\binits{P.~B.}},
\bauthor{\bsnm{Scalf},~\bfnm{Rusty}\binits{R.}},
\bauthor{\bsnm{Gunier},~\bfnm{Robert}\binits{R.}},
\bauthor{\bsnm{Smorodinsky},~\bfnm{Svetlana}\binits{S.}},
\bauthor{\bsnm{Wall},~\bfnm{Steve}\binits{S.}} \AND
\bauthor{\bsnm{Jerrett},~\bfnm{Michael}\binits{M.}}
(\byear{2006}).
\btitle{Nitrogen dioxide prediction in Southern California using land use
regression modeling: Potential for environmental health analyses}.
\bjournal{J. Expo. Sci. Environ. Epidemiol.}
\bvolume{16}
\bpages{106--114}.
\bid{doi={10.1038/sj.jea.7500442}, issn={1559-0631}, pii={7500442},
pmid={16047040}}
\bptok{imsref}%
\end{barticle}
%
\endbibitem

\bibitem[\protect\citeauthoryear{Ryan et~al.}{2005}]{Ryan2005}
%
\begin{barticle}[author]
\bauthor{\bsnm{Ryan},~\bfnm{P.~H.}\binits{P.~H.}},
\bauthor{\bsnm{LeMasters},~\bfnm{G.}\binits{G.}},
\bauthor{\bsnm{Biagini},~\bfnm{J.}\binits{J.}},
\bauthor{\bsnm{Bernstein},~\bfnm{D.}\binits{D.}},
\bauthor{\bsnm{Grinshpun},~\bfnm{S.~A.}\binits{S.~A.}},
\bauthor{\bsnm{Shukla},~\bfnm{R.}\binits{R.}},
\bauthor{\bsnm{Wilson},~\bfnm{K.}\binits{K.}},
\bauthor{\bsnm{Villareal},~\bfnm{M.}\binits{M.}},
\bauthor{\bsnm{Burkle},~\bfnm{J.}\binits{J.}} \AND
\bauthor{\bsnm{Lockey},~\bfnm{J.}\binits{J.}}
(\byear{2005}).
\btitle{Is it traffic type, volume, or distance? Wheezing in infants living
near truck and bus traffic}.
\bjournal{Journal of Allergy and Clinical Immunology}
\bvolume{116}
\bpages{279--284}.
\bptok{imsref}%
\end{barticle}
%
\endbibitem

\bibitem[\protect\citeauthoryear{Schikowski et~al.}{2005}]{Schikowski2005}
%
\begin{barticle}[author]
\bauthor{\bsnm{Schikowski},~\bfnm{T.}\binits{T.}},
\bauthor{\bsnm{Sugiri},~\bfnm{D.}\binits{D.}},
\bauthor{\bsnm{Ranft},~\bfnm{U.}\binits{U.}},
\bauthor{\bsnm{Gehring},~\bfnm{U.}\binits{U.}},
\bauthor{\bsnm{Heinrich},~\bfnm{J.}\binits{J.}},
\bauthor{\bsnm{Wichmann},~\bfnm{H.~E.}\binits{H.~E.}} \AND
\bauthor{\bsnm{Kramer},~\bfnm{U.}\binits{U.}}
(\byear{2005}).
\btitle{Long-term air pollution exposure and living close to busy roads are
associated with COPD in women}.
\bjournal{Respiratory Research}
\bvolume{6}
\bpages{152}.
\bptok{imsref}%
\end{barticle}
%
\endbibitem

\bibitem[\protect\citeauthoryear{Skene et~al.}{2010}]{Skene2010}
%
\begin{barticle}[pbm]
\bauthor{\bsnm{Skene},~\bfnm{Katherine~J.}\binits{K.~J.}},
\bauthor{\bsnm{Gent},~\bfnm{Janneane~F.}\binits{J.~F.}},
\bauthor{\bsnm{McKay},~\bfnm{Lisa~A.}\binits{L.~A.}},
\bauthor{\bsnm{Belanger},~\bfnm{Kathleen}\binits{K.}},
\bauthor{\bsnm{Leaderer},~\bfnm{Brian~P.}\binits{B.~P.}} \AND
\bauthor{\bsnm{Holford},~\bfnm{Theodore~R.}\binits{T.~R.}}
(\byear{2010}).
\btitle{Modeling effects of traffic and landscape characteristics on ambient
nitrogen dioxide levels in Connecticut}.
\bjournal{Atmos. Environ. (1994)}
\bvolume{44}
\bpages{5156--5164}.
\bid{doi={10.1016/j.atmosenv.2010.08.058}, issn={1352-2310}, mid={NIHMS237230},
pmcid={2976574}, pmid={21076636}}
\bptok{imsref}%
\end{barticle}
%
\endbibitem

\bibitem[\protect\citeauthoryear{Smith, Zhang and Field}{2007}]{Smith2007}
%
\begin{barticle}[mr]
\bauthor{\bsnm{Smith},~\bfnm{Brian~J.}\binits{B.~J.}},
\bauthor{\bsnm{Zhang},~\bfnm{Lixun}\binits{L.}} \AND
\bauthor{\bsnm{Field},~\bfnm{R.~William}\binits{R.~W.}}
(\byear{2007}).
\btitle{Iowa radon leukaemia study:~A hierarchical population risk
model for
spatially correlated exposure measured with error}.
\bjournal{Stat. Med.}
\bvolume{26}
\bpages{4619--4642}.
\bid{doi={10.1002/sim.2884}, issn={0277-6715}, mr={2411891}}
\bptok{imsref}%
\end{barticle}
%
\endbibitem

\bibitem[\protect\citeauthoryear{Son, Bell and Lee}{2011}]{Son2011}
%
\begin{barticle}[author]
\bauthor{\bsnm{Son},~\bfnm{J.}\binits{J.}},
\bauthor{\bsnm{Bell},~\bfnm{M.}\binits{M.}} \AND
\bauthor{\bsnm{Lee},~\bfnm{J.~T.}\binits{J.~T.}}
(\byear{2011}).
\btitle{Survival analysis to estimate the association between long-term
exposure to different sizes of airborne particulate matter and risk of infant
mortality using a birth cohort in Seoul, Korea}.
\bjournal{Epidemiology}
\bvolume{22}
\bpages{S166--S167}.
\bptok{imsref}%
\end{barticle}
%
\endbibitem

\bibitem[\protect\citeauthoryear{Venn et~al.}{2000}]{Venn2000}
%
\begin{barticle}[pbm]
\bauthor{\bsnm{Venn},~\bfnm{A.}\binits{A.}},
\bauthor{\bsnm{Lewis},~\bfnm{S.}\binits{S.}},
\bauthor{\bsnm{Cooper},~\bfnm{M.}\binits{M.}},
\bauthor{\bsnm{Hubbard},~\bfnm{R.}\binits{R.}},
\bauthor{\bsnm{Hill},~\bfnm{I.}\binits{I.}},
\bauthor{\bsnm{Boddy},~\bfnm{R.}\binits{R.}},
\bauthor{\bsnm{Bell},~\bfnm{M.}\binits{M.}} \AND
\bauthor{\bsnm{Britton},~\bfnm{J.}\binits{J.}}
(\byear{2000}).
\btitle{Local road traffic activity and the prevalence, severity, and
persistence of wheeze in school children: Combined cross sectional and
longitudinal study}.
\bjournal{Occup. Environ. Med.}
\bvolume{57}
\bpages{152--158}.
\bid{issn={1351-0711}, pmcid={1739915}, pmid={10810096}}
\bptok{imsref}%
\end{barticle}
%
\endbibitem

\bibitem[\protect\citeauthoryear{Wheeler et~al.}{2008}]{Wheeler2008}
%
\begin{barticle}[author]
\bauthor{\bsnm{Wheeler},~\bfnm{A.~J.}\binits{A.~J.}},
\bauthor{\bsnm{Smith-Doiron},~\bfnm{M.}\binits{M.}},
\bauthor{\bsnm{Xu},~\bfnm{X.}\binits{X.}},
\bauthor{\bsnm{Gilbert},~\bfnm{N.~L.}\binits{N.~L.}} \AND
\bauthor{\bsnm{Brook},~\bfnm{J.~R.}\binits{J.~R.}}
(\byear{2008}).
\btitle{Intra-urban variability of air pollution in Windsor,
Ontario---Measurement and modeling for human exposure assessment}.
\bjournal{Environ. Res.}
\bvolume{106}
\bpages{7--16}.
\bptok{imsref}%
\end{barticle}
%
\endbibitem

\bibitem[\protect\citeauthoryear{Zhang}{2011}]{Zhang2011}
%
\begin{bmisc}[author]
\bauthor{\bsnm{Zhang},~\bfnm{Lixun}\binits{L.}}
(\byear{2011}).
\bhowpublished{A Bayesian spatio-temporal model for estimating daily nitrogen
dioxide levels. Ph.D. thesis, Yale Univ., New Haven, CT}.
\bptok{imsref}%
\end{bmisc}
%
\endbibitem

\bibitem[\protect\citeauthoryear{Zmirou et~al.}{2002}]{Zmirou2002}
%
\begin{bmisc}[pbm]
\bauthor{\bsnm{Zmirou},~\bfnm{D.}\binits{D.}},
\bauthor{\bsnm{Gauvin},~\bfnm{S.}\binits{S.}},
\bauthor{\bsnm{Pin},~\bfnm{I.}\binits{I.}},
\bauthor{\bsnm{Momas},~\bfnm{I.}\binits{I.}},
\bauthor{\bsnm{Just},~\bfnm{J.}\binits{J.}},
\bauthor{\bsnm{Sahraoui},~\bfnm{F.}\binits{F.}},
\bauthor{\bsnm{Moullec},~\bfnm{Y.~Le}\binits{Y.~L.}},
\bauthor{\bsnm{Br{\'{e}}mont},~\bfnm{F.}\binits{F.}},
\bauthor{\bsnm{Cassadou},~\bfnm{S.}\binits{S.}},
\bauthor{\bsnm{Albertini},~\bfnm{M.}\binits{M.}},
\bauthor{\bsnm{Lauvergne},~\bfnm{N.}\binits{N.}},
\bauthor{\bsnm{Chiron},~\bfnm{M.}\binits{M.}},
\bauthor{\bsnm{Labb{\'{e}}},~\bfnm{A.}\binits{A.}} \AND
\borganization{VESTA Investigators}
(\byear{2002}).
\bhowpublished{Five epidemiological studies on transport and asthma: Objectives,
design and descriptive results.
\textit{J. Expo. Anal. Environ. Epidemiol.}
\textbf{12}
186--196}.
\bid{doi={10.1038/sj.jea.7500217}, issn={1053-4245}, pmid={12032815}}
\bptok{imsref}%
\end{bmisc}
%
\endbibitem

\end{thebibliography}
\end{document}